# Doping dependence of upper critical field and Hall resistivity in LaFeAsO$_{1-x}$F$_x$

Y. Kohama[1,2], Y. Kamihara[3,4], S. A. Baily[1,5], L. Civale[5], S. C. Riggs[6], F. F. Balakirev[1], T. Atake[2], M. Jaime[1], M. Hirano[3,4], and H. Hosono[3,4]

[1]*MPA-NHMFL, Los Alamos National Laboratory, Los Alamos, NM 87545, USA*

[2]*Materials and Structures Laboratory, Tokyo Institute of Technology, 4259 Nagatsuta-cho, Midori-ku, Yokohama 226-8503, Japan*

[3]*ERATO-SORST, Japan Science and Technology Agency in Frontier Research Center, Tokyo Institute of Technology, 4259 Nagatsuta-cho, Midori-ku, Yokohama 226-8503, Japan*

[4]*Frontier Research Center, Tokyo Institute of Technology, 4259 Nagatsuta-cho, Midori-ku, Yokohama 226-8503, Japan*

[5]*Superconductivity Technology Center, Los Alamos National Laboratory, Los Alamos, NM 87545, USA*

[6]*NHMFL, Florida State University, Tallahassee, FL, 32310, USA*

Abstract

The electrical resistivity ($\rho_{xx}$) and Hall resistivity ($\rho_{xy}$) of LaFeAsO$_{1-x}$F$_x$ have been measured over a wide fluorine doping range $0 \leq x \leq 0.14$ using 60 T pulsed magnets. While the superconducting phase diagram ($T_c$, $x$) displays the classic dome-shaped structure, we find that the resistive upper critical field ($H_{c2}$) increases monotonically with decreasing fluorine concentration, with the largest $H_{c2} \geq 75$ T for $x = 0.05$. This is reminiscent of the composition dependence in high-$T_c$ cuprates and might correlate with opening of a pseudo-gap in the underdoped region. Further, the temperature dependence of $H_{c2}(T)$ for superconducting samples can be understood in terms of multi-band superconductivity. $\rho_{xy}$ data for non-superconducting samples show non-linear field dependence, which is also consistent with a multi-carrier scenario.

The recent discovery of superconductivity above 30 K in a new class of iron-arsenide superconductors has attracted great interest [1]. Substantial numbers of theoretical and experimental studies have focused on striking similarities between iron-arsenides and high-$T_c$ cuprates such as the pseudogap phenomena [2] and magnetic order in the undoped parent compound [3]. However, some substantial differences have surfaced so far between iron-arsenides and the cuprates [4,5]. This includes proposals for multiband superconductivity [5,6], which is a characteristic feature of $MgB_2$ rather than high-$T_c$ cuprates. In this regard, there is yet no clear understanding of how the electron-doping affects the multiband superconductivity and electronic structure in iron-arsenides. In this report, we address the effects induced by fluorine-doping on transport properties in very high magnetic fields.

One of the properties of a superconductor that can be readily compared with theoretical models is the upper critical field ($H_{c2}$) which, although difficult to obtain directly, can be estimated from resistivity ($\rho_{xx}$) measurements. Indeed $H_{c2}(T)$, by comparing with models, can shed light on microscopic parameters such as the superconducting coherence length, superconducting gap and the superfluid density [7], and can determine whether this compound shows multiband superconductivity [8]. Another transport property related to the carrier number, that can be used to provide information about the multiband structure in the normal state of a superconductor, is the Hall resistivity ($\rho_{xy}$). Several recent experimental efforts on $LaFeAsO_{1-x}F_x$ have focused on $H_{c2}(T)$ and $\rho_{xy}$, but only in a limited range of magnetic fields and sample compositions [5,9,10]. Hence a consistent interpretation of results and identification of the relevant mechanisms are still missing. In this work, we present an effort to understand high-magnetic-field $\rho_{xx}$ and $\rho_{xy}$ in $LaFeAsO_{1-x}F_x$ samples with a broad range of compositions, synthesized by our group, and address the fundamental question of how $H_{c2}$ and $\rho_{xy}$ of $LaFeAsO_{1-x}F_x$ are affected by F-doping.

Polycrystalline samples of LaFeAsO$_{1-x}$F$_x$ ($x$ = 0, 0.025, 0.05, 0.07, 0.11 and 0.14) were prepared by solid-state reactions, as described elsewhere [1]. The phase purity was checked by powder X-ray diffraction (XRD) using Cu-K alpha source (Bruker D8 Advance TXS) and synchrotron radiation source at SPring-8 ($x$ = 0 and 0.14). All the XRD patterns did not show any peak separations indicating the high sample quality. For $x \geq 0.025$, the magnetic field dependence of the $\rho_{xx}$ and $\rho_{xy}$ were measured at fixed temperatures using a capacitor bank-driven 60 T pulsed magnet. Both properties were measured simultaneously using a lockin detection technique operating at 80-102 kHz in a standard 6-contact Hall bar configuration. For $x$ = 0, $\rho_{xy}$ was measured using a Quantum Design® Physical Properties Measurement System. Measurements of the magnetization for $x$ = 0.05 were performed using a Quantum Design® Magnetic Properties Measurement System.

Figure 1 shows $\rho_{xx}$ versus magnetic field ($H$) for superconducting samples $x$ = 0.05, 0.07, 0.11 and 0.14. The $\rho_{xx}(H)$ curves for $x$ = 0.14 have a narrow field range of zero resistivity and show a double kink structure not seen for the other compositions. In Fig. 2, $\rho_{xx}(T)$ also shows a bit of structure, raising the concern of sample homogeneity close to the superconductor-normal metal boundary [1]. However, our XRD data does not show signs of phase segregation, and thus we believe that the structure in $\rho_{xx}(H)$ and $\rho_{xx}(T)$ is intrinsic and is the signature of multiband superconductivity altering the anisotropy and the influence of grain boundaries, respectively. Figure 1 shows that all samples display positive magnetoresistance $\Delta\rho_{xx}(H) = \rho_{xx}(H) - \rho_{xx}(0)$ in the superconducting state as well as in the normal state for temperatures close to the superconducting transition temperature ($T_c$). The estimated $H_{c2}$ for underdoped ($x$ = 0.05 and 0.07) and optimally doped ($x$ = 0.11) samples are quite large, and the magnetic field can only reestablish the normal metallic state in a limited temperature (above ~10 K) and magnetic fields range (above ~50 T). We observe that $\rho_{xx}$ in the normal-state increases with decreasing temperature for $x$ = 0.05 (insulating behavior), while in other samples $\rho_{xx}$ decreases with decreasing temperature (metallic behavior). This is reminiscent of the insulator-to-metal crossover (IMC) induced by doping in the high-$T_c$ cuprates [11,12], although in contrast with the high-$T_c$ cuprates the parent compound LaFeAsO displays finite conductivity at $T$ = 0 [1,13]. In order to examine the metal-insulator crossover more closely, we plot $\rho_{xx}$ under various magnetic fields as a function of temperature in Fig. 2. This way of displaying the data confirms that the $x$ = 0.05 sample exhibits a clear upturn at low temperatures, while the $\rho_{xx}$ for $x \geq 0.07$ decreases with decreasing temperature. A similar resistive upturn was reported in the low doping members of the SmFeAsO$_{1-x}$F$_x$ system [14].

One of the most important parameters that can be extracted from the $\rho_{xx}(H)$ data is the upper critical field ($H_{c2}$). To accomplish this in a consistent way across the compositions, we estimate the zero field normal-state resistivity ($\rho_{xx}^n$) with fits of the type $\rho_{xx}^n = a + bT + cT^2$ and $\rho_{xx}^n = d + eT^{-1}$ for $x \geq 0.07$ and $x$ = 0.05 (fig. 2), respectively. We add the temperature independent magnetoresistance term $\Delta\rho_{xx}(H)$ taken from the $\rho_{xx}(H)$ data near $T_c$, and evaluate the normal-state resistivity under magnetic field as $\rho_{xx}^n + \Delta\rho_{xx}(H)$. The upper critical fields ($H_{c2}^{80}$) are defined as the field value at which the measured resistivity is 80% of the $\rho_{xx}^n + \Delta\rho_{xx}(H)$. Similar procedures for determining $H_{c2}$ have been applied to a variety of superconducting systems [5, 15], and the $H_{c2}^{80}$ approximates the larger in-plane upper critical field ($H_{c2}^{\parallel}$) in polycrystalline samples. Here we note that the out-plane upper critical filed ($H_{c2}^{\perp}$) is difficult to evaluate from our polycrystalline data due to the irreversibility field.

Figure 3 shows the temperature dependence of $H_{c2}^{80}(T)$ evaluated for all of our samples. As clearly shown in this figure, $H_{c2}^{80}(T)$ systematically decreases with increasing $x$, and the shape of the $H_{c2}^{80}(T)$ curves also change. The $H_{c2}^{80}(0)$ are uncorrelated with $T_c$, and all our samples show ratios of $H_{c2}$ to $T_c$ (from 2.0 -2.3 T/K for x = 0.14 to 3.5 - 5.7 T/K for x = 0.05) that exceed the Pauli limit $H_P = 1.84\ T_c$. If we instead use the 90% and 50 % values of $\rho_{xx}^n + \Delta\rho_{xx}(H)$ for evaluation of $H_{c2}$, the overall tendency of $H_{c2}(0)$ does not change. A systematic decrease of $H_{c2}(0)$ with decreasing $x$ was reported in high-$T_c$ cuprates [16], and is considered as evidence for bosonic pairs that form above $T_c$ in the so-called "pseudogap" state, although the pairs are too dilute to form superconducting condensate. In the high-$T_c$ cuprates, this scenario was associated with a pseudogap which increases with decreasing $x$ [16]. Recently, a photoemission spectroscopy study [17] showed that the pseudogap in the LaFeAsO$_{1-x}$F$_x$ system also has similar $x$ dependence to that of high-$T_c$ cuprates. Perhaps the apparent increase of $H_{c2}(0)$ in underdoped LaFeAsO, which might correlate with the pseudogap opening, indicates pair formation mechanics similar to underdoped cuprates. On the other hand, the $H_{c2}^{80}(T)$ curve for $x = 0.14$ rapidly increases as $T \to 0$, which is similar to the $H_{c2}$ dependence in the multiband superconductor [18]. Recent theoretical and experimental studies pointed out the multiband nature of iron-pnictides [5,6], thus we try to fit the 2-band theoretical curve to our $H_{c2}$ data [8]; $a_0[\ln t + U(h)][\ln t + U(\eta h)] + a_1[\ln t + U(h)] + a_2[\ln t + U(\eta h)] = 0$ (Eq. 1). The constants $a_1$, $a_2$ and $a_3$ are determined from the BCS coupling constant tensor $\lambda_{mm'}$, and the other parameters are defined as $U(x) = \psi(1/2 + x) - \psi(1/2)$, $h = H_{c2}D_1/2\phi_0 T$, $t = T/T_c$ and $\eta = D_2/D_1$, where $\psi$ is the di-gamma function, $\phi_0$ is the magnetic flux quantum and $D_m$ is the electronic diffusivity for the $m^{th}$ Fermi surface sheet.

We use the inter-band coupling values ($\lambda_{12} = \lambda_{21} = 0.5$) from ref [5], and also take the same intra-band coupling values $\lambda_{11} = \lambda_{22} = 0.5$. The eq. 1 is then left with only two independent parameters $H_{c2}(0)$ and $\eta$. The solid curves in Fig.3 are the fits obtained using Eq.1, and the above defined fitting parameters are plotted in Fig. 4(a). The high quality of the fits supports the relevance of the two-band model for the LaFeAsO system. Despite significant error bars for $x = 0.05$, originating from its high $H_{c2}$ and the upturn in $\rho_{xx}^n$, the diffusivity ratio ($\eta$) shows a decrease of one order of magnitude with increasing $x$. When $\eta$ is equal to 1, the fitting curve corresponds to the traditional Werthamer-Helfand-Hohenberg (WWH) curve [8]. Thus the small $\eta$ values for $x = 0.11$ and 0.14 indicate that the traditional WHH fitting cannot reproduce our data. The small $\eta$ also means one band is dirtier than the other band, which reflects the change of the characteristic shape of $H_{c2}(T)$ with increasing $x$ [8]. In particular, the shape for $x = 0.14$ is similar to a previous report on carbon-doped $MgB_2$, which shows a much more rapid increase of $H_{c2}$ near $T = 0$ for carbon-doped $MgB_2$ than un-doped $MgB_2$ [18]. Since the fluorine dopant may act as a scattering center similar to the carbon dopant in $MgB_2$ [19], the characteristic shape of $H_{c2}(T)$ for $x \geq 0.11$ could originate from the enhancement of scattering in one of the bands.

The inset of Fig. 3 shows the field dependence of the dc magnetization for $x = 0.05$. The magnetization curves for $T = 1.8\text{-}18$ K are typical for a type-II superconductor and permit us to determine the lower critical field ($H_{c1}$). $H_{c1}$ is defined as the magnetic field where vortices enter the sample causing a departure from the linear behavior in magnetic moment vs. field. The filled squares in Fig. 3 represent the $H_{c1}$ and roughly show linear temperature dependence. Similar behavior has been observed in the $MgB_2$ [20,21], and it may originate from the multiband superconductivity in F-doped LaFeAsO. From the linear fit, we extrapolate the $H_{c1}$ at zero temperature $H_{c1}(0) = 6.0 \pm 0.5$ mT. With the estimation of coherence length ($\xi$) from $H_{c2}$, we may evaluate the penetration depth ($\lambda_L$) and the Ginzburg-Landau parameter ($\kappa$) using following equations, $H_{c2}(0) = \phi_0/2\pi\xi^2$, $H_{c1}(0) = (\phi_0/4\pi\lambda_L^2)\ln(\kappa)$ and $\kappa = \lambda_L/\xi$. These yield $\xi = 15 - 19$ Å, $\lambda_L = 3600 - 4100$ Å and $\kappa = 190 - 270$.

Our Hall coefficient studies in the normal state also reveal a behavior that is characteristic of a multiband electronic structure. Figures 5(a) and 5(b) display the Hall resistivity ($\rho_{xy}$) for $x = 0$ and 0.025, respectively, which show non-linear field dependence below ~150 K in sharp contrast to the expected linear response, $\rho_{xy} = R_H H$, in a single band metal. We could not detect any non-linear behavior for $x \geq 0.05$, although an observation of non-linear behavior for $x \geq 0.11$ becomes experimentally difficult due to the small value of $\rho_{xy}$. The non-linear behavior can be fit satisfactorily by the 2-band Drude model (fig. 5(a)) [22],

$$\rho_{xy}(H) = \frac{\sigma_h^2 R_h + \sigma_e^2 R_e + \sigma_h^2 \sigma_e^2 R_h R_e (R_h + R_e) H^2}{(\sigma_h + \sigma_e)^2 + \sigma_h^2 \sigma_e^2 (R_h + R_e)^2 H^2} H$$ (Eq. 2), where $\sigma_{e(h)}$ and $R_{e(h)}$ are the

electrical conductivity and the Hall coefficient of the electron (hole) band. However, there are too many independent variables to find a unique set of values for $\sigma_{e(h)}$ and $R_{e(h)}$. Equation 2 also predicts that the magnitude of the non-linear response is roughly proportional to the square of the ratio $\rho_{xy}/\rho_{xx}$ [22]. Therefore, a correlation of $\rho_{xy}/\rho_{xx}$ with the non-linear response would provide additional evidence for multiband electronic band structure. Thus, we fit our data to the polynomial equation $\rho_{xy} = R_H H + \beta H^3$ in order to estimate the magnitude of the linear ($R_H$) and non-linear term ($\beta$). Displayed in Fig. 5(c), the non-linear response $\beta$ rapidly decreases with increasing temperature and F-dopant, and vanishes above 150 K. As seen in the inset of Fig. 5(d), $-\rho_{xy}/\rho_{xx}$ also show a rapid decrease with the rise of temperature and increasing F-dopant. This clear correlation between $-\rho_{xy}/\rho_{xx}$ and $\beta$ implies that the disappearance of non-linear behavior is also consistent with the 2-band Drude model. However, we note that the non-linear behavior might also be due to the change of the carrier number, for example by partial closing of the gap under magnetic field.

The linear contribution ($R_H$) corresponds to the low-field limit of the 2-band Drude model, which is given by the following equation:

$$R_H = \frac{\sigma_h^2 R_h}{(\sigma_h + \sigma_e)^2} + \frac{\sigma_e^2 R_e}{(\sigma_h + \sigma_e)^2} \quad \text{(Eq.3)},$$

where $R_{e(h)}$ relate to the carrier density at each band ($n_{e(h)}$) and the carrier charge ($q$) as $R_{e(h)} = 1/n_{e(h)}q$. Figure 5(d) shows the temperature dependence of $|R_H|$ for the entire composition range (which agrees with recent results for $x = 0.11$ [9]), where the sign of $R_H$ for all samples is negative. $|R_H|$ systematically decreases with increasing $x$, also shown in fig. 4 (b), and increases with decreasing $T$. The non-trivial temperature dependence of $|R_H|$ might be explained by the contributions from electron and hole bands in Eq.3 having different temperature dependences. On the other hand, it is hard to explain the $x$ dependence of $|R_H|$ using Eq. 3. Since the F-doping should act as electron doping, the $x$ dependence could originate from the decrease of $|R_e|$. The most pronounced features in fig. 5(d) are the rapid upturns of $|R_H|$ observed for $x = 0$ and 0.025 below 150 K. These seem to relate to the structural and/or magnetic phase transition observed by other measurements [3,13]. In fact, as shown in Fig. 4(a), the temperature ($T_g$) detected by $R_H$ is similar to the temperature at which the susceptibility shows an anomaly ($T_A$) [13]. One plausible scenario explaining the rapid upturns is the decrease of $n_e$ suggesting opening of an energy gap at the Fermi level, which is in agreement with the recent optical spectroscopy results in similar compound [23].

In Fig. 4, we summarize the parameters determined in this work, together with the density of states ($N_D$) estimated by heat capacity and magnetic susceptibility [13]. While $T_c$ and $N_D$ doping dependence is dome shaped, $R_H$ and $H_{c2}$ change monotonically with $x$. If the inverse of $R_H$ is proportional to the actual carrier number ($n$) which is expected in a single band model, the bare density of states ($N_D^{cal}$) cannot display a peak structure in the parabolic band limit ($N_D^{cal} \propto n^{1/3}$). Since strong spin fluctuations were reported [13], the peak of $N_D$ may be induced by a spin-fluctuation-related renormalization. The alternative, i.e. the peak structure in $N_D^{cal}(x)$ is true, seems to be inconsistent with band calculations [24]. On the other hand, we want to point out that the doping dependence of $H_{c2}$ might be understood in terms of the increment of $N_D$, because the $N_D$ can be a factor to enhance the Pauli limit $H_P = 1.84\ T_c$, $H_P^* = H_P(1 + VN_D)^{0.5}$ where $V$ is the average matrix element used by BCS theory [25]. For more detailed discussion, more precise measurements are needed, which will be produced by systematic measurements on single crystals.

We have shown the effect of F-doping on $\rho_{xx}$ and $\rho_{xy}$ from the undoped to the overdoped region.

The $\rho_{xx}(H)$ curves allow for an estimation of $H_{c2}(T)$ throughout the entire $x$ range, and reveal an increase of $H_{c2}(0)$ with decreasing $x$ down to $x = 0.05$. The similar increase of $H_{c2}$ was also reported in high-$T_c$ cuprates, where it was proposed that the increase of $H_{c2}$ results from an increase of superconducting pairing potential with decreasing $x$. In addition, we can fit the curves of $H_{c2}(T)$ in the entire composition region to multi-band model. We found the non-linear behavior in $\rho_{xy}$ for non-superconducting samples, which also provide an evidence of the multi-carrier system. The non-linear $\rho_{xy}$ could be a key to the understanding of the electronic structure in the iron-arsenides.

Acknowledgements

This work was supported by a Grant-in-Aid JSPS (Grant No. 19.9728), the NSF Division of Materials Research through DMR-0654118, the U.S. Department of Energy and the State of Florida. We would like to thank H. Kawaji, S. Francoual and H. Yuan for technical assistance during the experiment.

Figure captions

Figure 1 (color) (a)-(d): Magnetic field dependence of resistivity $\rho_{xx}(H)$ for $x = 0.05$ (a), $x = 0.07$ (b), $x = 0.11$ (c) and $x = 0.14$ (d).

Figure 2 (color) (a)-(d): Temperature dependence of resistivity $\rho_{xx}(T)$ for $x = 0.05, 0.07, 0.11$ and $0.14$. The black dots are $\rho_{xx}(T)$ at zero field. The colored dots are extracted from $\rho_{xx}(H)$. The dashed line indicates the normal state resistivity at 0 T (see text).

Figure 3 (color) Main Panel: $H_{c2}^{80}$ versus temperature. The open circles correspond to $H_{c2}^{80}$ evaluated by the extrapolation of $\rho_{xx}(H)$ in Fig. 1. The solid curves represent $H_{c2}(T)$ calculated from the two-band theory [8]. The yield parameters ($\eta$ and $H_{c2}(0)$) are shown in the Fig. 4(a). Inset: Magnetic-field dependence of the zero-field cooled magnetization at various temperatures 1.8 – 18 K. The arrow indicates $H_{c1}$ at 1.8 K.

Figure 4 (color) Summary of the doping dependent properties in LaFeAsO$_{1-x}$F$_x$. $\eta$, $H_{c2}(0)$, $T_c$, $T_g$ and $T_A$ are shown in Fig. 4(a). $R_H$ at 50 K is plotted together with the density of states calculated from heat capacity ($N_D^{\gamma}$) and magnetic susceptibility ($N_D^{\chi}$) in Fig. 4(b)[13].

Figure 5 (color) (a)(b): Magnetic field dependence of the Hall resistance $\rho_{xy}(H)$ for $x = 0$ and 0.025. Solid squares in Fig. 5(a) and solid curves in Fig. 5 (b) represent data taken in static magnetic fields and pulsed field. The solid curves in Fig. 5(a) are the results of the 2-band fit. The data in Fig. 5(b) can also be fit by a 2-band model (not shown). (c): Coefficient of $\beta$ in the fit of $\rho_{xy}(H) = R_H H + \beta H^3$. Here the data points for $x = 0.025$ are multiplied by 50. (d): Temperature dependence of $R_H$. The dashed curve is taken from Ref. 9. The inset shows the ratio of the -$\rho_{xy}(10\ T)$ to $\rho_{xx}$ at 50, 100 and 200 K.

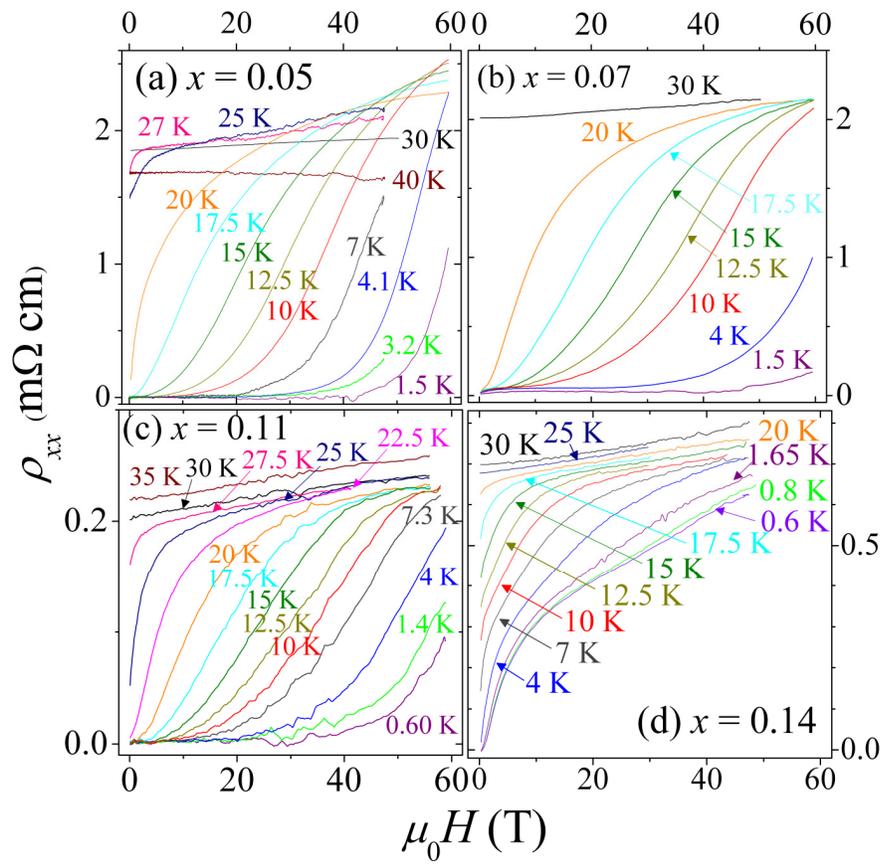

Fig. 1.   Y. KOHAMA, et al.

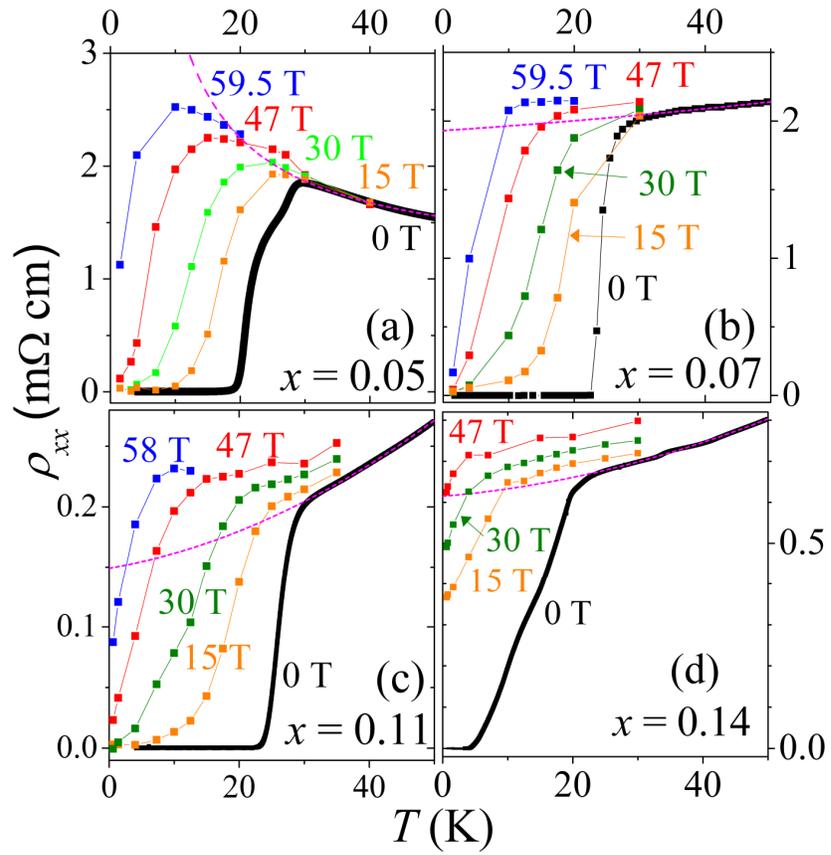

Fig. 2.   Y. KOHAMA, et al.

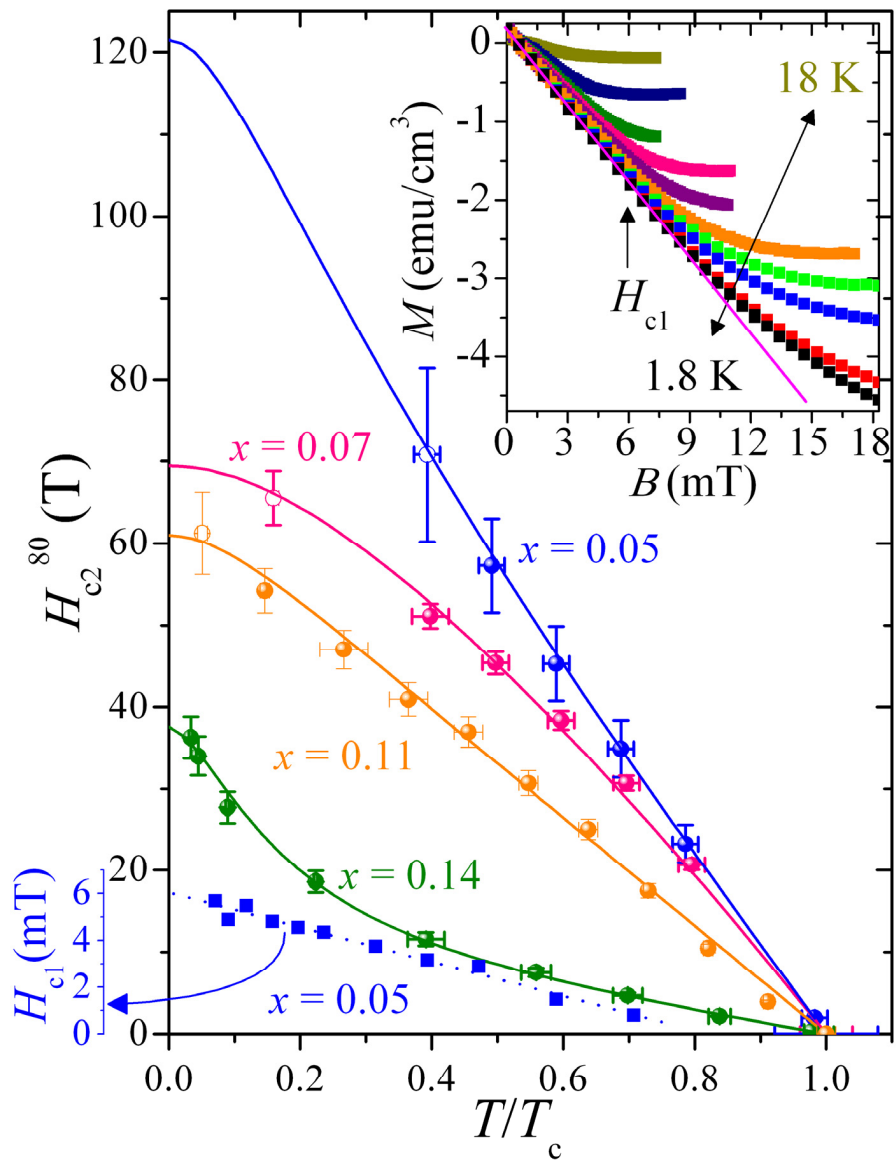

Fig. 3.   Y. KOHAMA, et al.

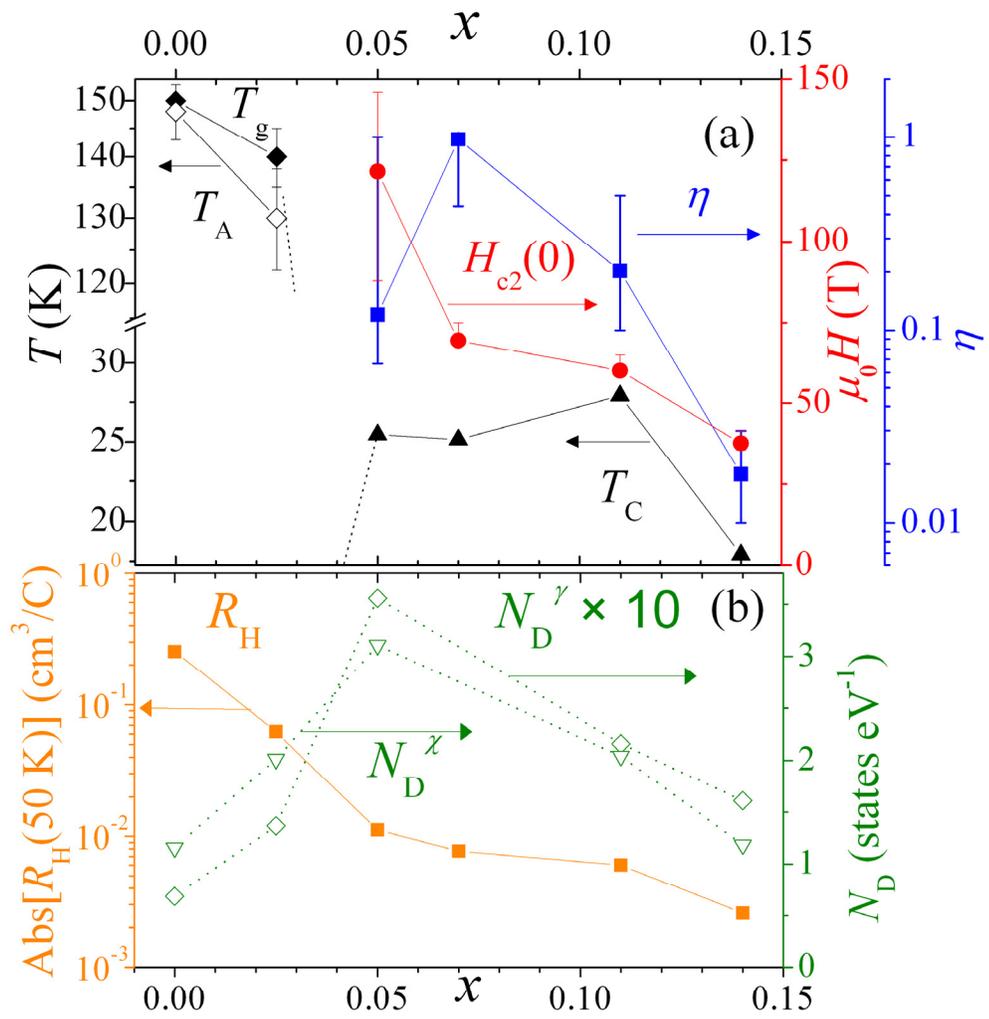

Fig. 4.    Y. KOHAMA, et al.

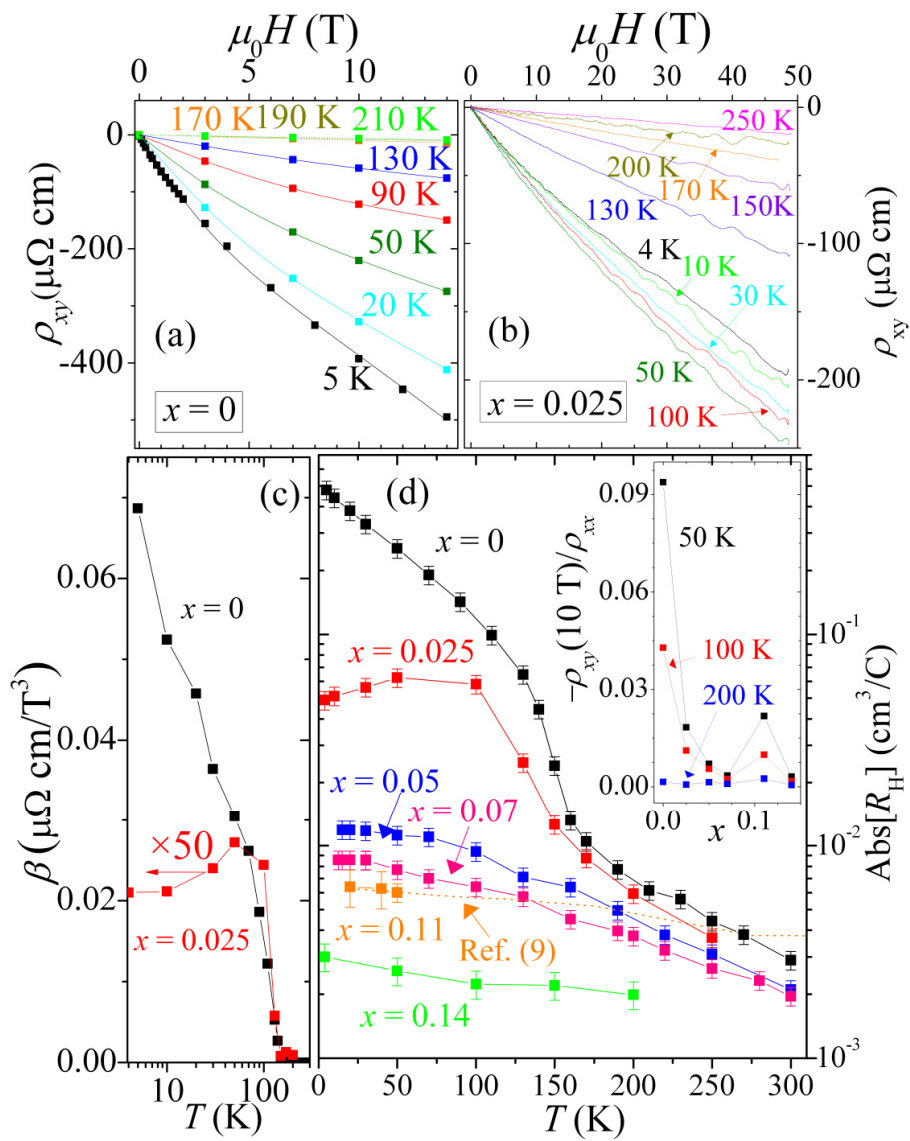

Fig. 5.    Y. KOHAMA, et al.